\documentclass[12pt]{article}
\usepackage{graphicx}
\usepackage{amssymb}
\usepackage{amsfonts}
\usepackage{amsthm}
\usepackage{amsfonts}

\newcommand \coord{{z}}

\newcommand \half {{\mathbb H}}

\newcommand{\integers}{{\mathbb Z}}

\newcommand \verbose {0}

\newcommand \no {\noindent}
\newcommand \ra {\rightarrow}

\newcommand{\ba}[1]{\begin{array}{#1}}
\newcommand{\ea}{\end{array}}

\newcommand{\be}{\begin{equation}}
\newcommand{\ee}{\end{equation}}
\newcommand{\bea}{\begin{eqnarray}}
\newcommand{\eea}{\end{eqnarray}}
\newcommand{\beann}{\begin{eqnarray*}}
\newcommand{\eeann}{\end{eqnarray*}}

\newcommand{\map}{{f}}
\newcommand{\walk}{{\overline{\omega}}}
\newcommand{\Z}{{\cal Z}}
\newcommand{\varn}{{n}}
\newcommand{\fixn}{{N}}

\def\reff#1{(\ref{#1})}
\begin{document}

\bibstyle{ams}

\title{Conformal invariance predictions for the \\
three-dimensional self-avoiding walk
}

\author{Tom Kennedy
\\Department of Mathematics
\\University of Arizona
\\Tucson, AZ 85721
\\ email: tgk@math.arizona.edu
}

\maketitle 

\begin{abstract}
If the three dimensional self-avoiding walk (SAW) is 
conformally invariant, then one can compute the hitting densities 
for the SAW in a half-space and in a sphere \cite{kennedy_3d_prl}.
The ensembles of SAW's used to define these hitting densities 
involve walks of arbitrary lengths, and so these ensembles cannot
be directly studied by the pivot Monte Carlo algorithm for the SAW.
We show that these mixed length ensembles should have the same 
scaling limit as certain weighted ensembles that only involve
walks with a single length, thus providing a fast method for 
simulating these ensembles.
Preliminary simulations which found good agreement
between the predictions and Monte Carlo simulations for the SAW 
were reported in  \cite{kennedy_3d_prl}.
In this paper we present more accurate 
simulations testing the predictions and find even stronger support
for the prediction that the SAW is conformally invariant in 
three dimensions.
\end{abstract}

\newpage

\section{Introduction}

The scaling limit of the self-avoiding walk (SAW) is 
conjectured to be conformally invariant. In two dimensions this
conjecture leads to a complete description of the scaling limit.
If $D$ is a simply connected domain, the scaling limit of the 
SAW between two fixed boundary points is conjectured to 
be chordal Schramm-Lowener evolution (SLE)
with parameter $\kappa=8/3$ \cite{lsw_saw}. 
(Between a fixed boundary point and a fixed interior point 
the scaling limit is conjectured to be radial SLE.) 
If one considers SAW's in $D$ which start at a fixed interior
point but are allowed to end anywhere on the boundary, 
then there are predictions for the distribution of the terminal point 
\cite{lsw_saw}.
Monte Carlo tests of the SLE predictions for the two-dimensional
SAW can be found in \cite{dyhr_et_al,Kennedya,Kennedyb}.
Tests of the predictions for the hitting density can be found in 
\cite{kennedy_lawler,kennedy_dilation}
All these simulations found excellent agreement with 
the predictions, supporting the conjecture that the two-dimensional
SAW is conformally invariant.  

The success of using conformal invariance to study the 
two dimensional SAW is a result of the large group of 
conformal transformations in two dimensions. 
In three and higher dimensions this group consists only of 
compositions of Euclidean symmetries and inversions in spheres. 
It is not expected that invariance under this limited group 
completely determines the SAW in three dimensions, 
but one can still ask if the SAW is 
conformally invariant and if this invariance leads to any non-trivial 
predictions for the three-dimensional SAW.
We restrict our attention to three dimensions since the 
scaling limit of the SAW has been proved to be Brownian motion 
in more than four dimensions \cite{hara_slade_a,hara_slade_b}, and 
substantial progress toward proving this in four dimensions 
has been made \cite{bs}.

In \cite{kennedy_3d_prl} we argued that the conformal covariance
prediction of \cite{lsw_saw} for the hitting density 
in two dimensions holds in three dimensions as well. 
This led to a prediction of the hitting
density for the SAW starting at an interior point of 
the half-space $z>0$ and the hitting density of the SAW 
starting at an interior point of a sphere. 
The conjectured conformal invariance also gives a prediction for
the distribution of the first time the SAW in the full-space 
between two fixed points crosses the plane which bisects
the line segment between those two points. 
These three predictions for the SAW were tested by Monte Carlo 
simulations of the SAW, and the results of those simulations 
were reported in \cite{kennedy_3d_prl}.
We found good agreement between the simulations and the conformal 
invariance predictions, supporting the conjecture that the 
SAW is conformally invariant in three dimensions. 

The ensembles of SAW's in these predictions are not easily studied
by the usual Monte Carlo algorithms for the SAW, in particular
the pivot algorithm. The simulations presented in \cite{kennedy_3d_prl}
relied on a representation of these ensembles in terms of the 
ensemble of SAW's with a fixed number of steps and no constraint
on the end point. In this paper we derive these representations.
We also explain the lattice effects that persist in the scaling 
limit that must be taken into account. 

In the next section we review the definition of the SAW and the predictions 
derived in \cite{kennedy_3d_prl}. 
In section three we derive the representations of the ensembles we 
wish to study in terms of the ensemble of SAW's with a fixed number
of steps. The lattice effects that persist in the scaling limit are 
studied in section four. 
Section five presents the results of the simulations, and the final 
section presents our conclusions.

\section{Definitions and predictions}

We review the definition of the self-avoiding walk. 
We introduce a lattice with spacing $\delta$, e.g., $\delta \integers^d$.
Our simulations use the cubic lattice, but our 
predictions about the scaling limit should hold for any regular lattice
in three dimensions. 
A self-avoiding walk is a nearest-neighbor 
walk on the lattice which does not visit any site in the lattice more
than once. We will denote such a walk by $\omega(i)$ 
where $i=0,1,2,\cdots,N$  with $N$ being the number of steps in the walk. 
So $\omega$ satisfies $|\omega(i)-\omega(i-1)|=\delta$ 
for $i=1,2,\cdots,N$ and 
$\omega(i) \neq \omega(j)$ for $i \neq j$. In this paper we will 
always take the SAW to start at the origin, i.e., $\omega(0)=0$. 
There are a variety of ensembles that one can consider. 
One can consider the SAW on the full lattice, or one can only allow 
SAW's that lie inside some prescribed domain. 
One can consider SAW's which start and end at prescribed points, or 
one can consider SAW's which start at a prescribed point but are allowed
to end in some set, e.g., the boundary of a domain. 

We start with the SAW in a bounded domain. 
Let $D$ be a bounded 
domain containing the origin or with the origin on its boundary. 
The ordinary random walk in $D$ started at the origin and stopped when it 
exits $D$ may be described as follows. We consider all finite 
nearest neighbor walks in $D$ which start at the origin and stay inside 
$D$ except for the last step of the walk which takes it outside $D$. 
Such a walk $\omega$ is given probability $\coord^{-|\omega|}$ 
where $\coord$ is the coordination number of the lattice. 
We define the SAW in $D$ starting at the origin in an analogous
fashion. In place of the coordination number we use the connective 
constant $\mu$. It is given by $\mu = \lim_{N \ra \infty} c_N^{1/N}$ where 
$c_N$ denotes the number of SAW's on the full lattice 
with $N$ steps that start at $0$. 
(The existence of this limit follows from the  
subadditivity of $\ln(c_N)$ \cite{madras_slade}.)
The SAW in $D$ is then defined by taking all SAW's that start at the 
origin and stay inside $D$ except for the last bond which takes the walk 
outside of $D$. The probability of such a walk $\omega$ is defined 
to be proportional to $\mu^{-|\omega|}$. (The constant of proportionality
is determined by the requirement that we get a probability measure.) 
Rather than allow the SAW to end at any site just outside the boundary, 
we could also fix both the starting point and an ending point 
just outside $D$ and consider the ensemble of SAW's between these two 
points that stay inside $D$ except for the last step.
When $D$ is unbounded, there 
are infinitely many SAW's between the two points, and it is not 
clear whether the sum of the weight $\mu^{-|\omega|}$ is finite. 

We would also like to consider the SAW in an unbounded domain when
the terminal point is taken to be $\infty$. In this case we 
need to modify the definition. There are two possible definitions.
For the first, we fix an integer $N$ and take all SAW's 
which start at the prescribed starting point, stay inside $D$
and have exactly $N$ steps. 
We put the uniform probability measure on this finite set of walks. 
It is expected that as $N \ra \infty$, this probability measure 
converges weakly to a probability measure supported on infinite SAW's from the 
starting point to $\infty$. The other definition is to consider all
finite length SAW's starting at the prescribed starting point 
and staying in $D$. The weight of a walk $\omega$ is taken to 
be $c^{-|\omega|}$ with $c>\mu$. This inequality insures that the 
resulting measure is finite and so can be normalized to give a 
probability measure. We then expect that as $c \ra \mu^+$ this probability 
measure converges weakly to a probability 
measure supported on infinite SAW's from the 
starting point to $\infty$. It is expected that the two definitions 
produce the same probability measure. 

For all these ensembles the scaling limit should be obtained by 
letting $\delta \ra 0$. It is expected that the probability measures
converge weakly in this limit to probability measures supported on 
simple curves, but this has not been proven. 

We review the definitions of several critical exponents that we will use.
The average distance that an $N$ step SAW travels is described by 
the exponent $\nu$. 
\beann
E_N [||\omega(N)||^2] \sim N^{2 \nu}
\eeann
where $E_N$ is expectation with respect to the uniform probability 
measure on SAW's with $N$ steps starting at the origin. 
The exponents $\gamma$ and $\rho$ characterize the asymptotic 
behavior of the number of SAW's. 
The number of SAW's with $N$ steps starting at the origin
grows as $\mu^N N^{\gamma-1}$. 
If you only allow SAW's that stay in a half-space, e.g.,
$\{(x,y,z): z>0\}$, or in two dimensions in a half-plane, 
then it grows like $\mu^N N^{\gamma-1-\rho}$.
So the probability that a SAW with $N$ steps in the full-space
lies in the half-space goes like $N^{-\rho}$.
The final exponent we need is the boundary scaling exponent $b$.
The partition function of the SAW between two fixed boundary 
points of a domain goes as $\delta^{2b}$ as the lattice 
spacing $\delta$ goes to zero. The exponent $b$ also characterizes
the probability that a two-dimensional SAW will pass through a small slit 
in a curve. If $\epsilon$ is the width of the slit, the probability 
goes as $\epsilon^{2b}$ as $\epsilon$ goes to zero. This holds in
three dimensions as well if $\epsilon$ is the linear size of 
a small hole in a surface.

In two dimensions there are predictions for the values of these
exponents, none of which have been proved. (In fact the existence of 
these exponents is not even established rigorously in two and three
dimensions.) The predictions are
$\nu=3/4$  \cite{flory}, $\gamma=43/32$ \cite{nienhuis}, 
$\rho=25/64$ \cite{cardy1984conformal} and $b=\eta_\parallel/2=5/8$ 
\cite{cardy1984conformal}.
In three dimensions there are numerical estimates for these
exponents but no exact predictions. 
In three dimensions Clisby \cite{clisby_nu} finds  $\nu=0.587597(7)$
Schram, Barkema and Bisseling \cite{gamma_schram}, 
find $\gamma=1.15698(34)$.
Grassberger \cite{grassberger} finds 
$\gamma_1=0.6786 \pm 0.0012$ where  $\gamma_1=\gamma-\rho$.

These exponents are related by a standard scaling relation:
\bea
b =  {2\rho-\gamma \over 2 \nu} + {d \over 2} 
\label{bformula}
\eea
A heuristic derivation of this relation for the SAW can be found
in \cite{lsw_saw}. They gave the derivation 
in two dimensions, but the derivation works in any number of dimensions. 
\ifodd \verbose
Look at a square/cube of size $L$ and take the lattice spacing to be $1$. 
Fix two boundary points at centers of opposite edges/faces, call them 
$a$ and $b$. We look at 
\beann
\sum_{\omega:a \ra b} \mu^{-|\omega|}
\eeann
It should decay like $L^{-2b}$. (This is def of $b$.)
Rewrite as 
\beann
\sum_N \, \mu^{-N} \, \sum_{\omega:|\omega|=N, a \ra b} =\sum_N \mu^{-N} \, a_N
\eeann
where $a_N$ is number of $N$ step walks from $a$ to $b$ in the square/cube. 
$N$ must be such that $N^\nu$ is of order $L$. So the sum on $N$ has 
limits like $c_1 L^{1/\nu}$ to $c_2 L^{1/\nu}$.
So this sum gives a factor of $L^{1/\nu}$. The number of $N$ step walks 
starting at $a$ and staying on the correct side of the face containing 
$a$ is of order $\mu^N N^{\gamma-1-\rho}$. The number of sites where these 
walks can end is of order $L^d$, but we only want the ones ending at $b$. 
So we get a factor of $L^{-d}$. Finally, the walks must stay on the correct
side of the face containing $b$ which gives another factor of $N^{-\rho}$. 
So we have 
\beann
L^{1/\nu} N^{\gamma-1-\rho} L^{-d} N^{-\rho}
= L^{1/\nu - d +(\gamma-1-2\rho)/\nu}
\eeann
So we get 
\beann
-2b = {1 \over \nu}  - d +{\gamma-1-2\rho \over \nu}
\eeann
\else { } \fi

For an ensemble in which the terminal point is allowed to be anywhere
along the boundary we will refer to the distribution of this 
random terminal point as a hitting distribution. 
For the ordinary random walk the scaling limit of this distribution 
is harmonic measure. For the SAW there is another 
distribution on the boundary that can be defined. 
One can consider the SAW in the full-plane or space from 
$0$ to $\infty$ and look at the first exit from $D$. We emphasize that 
this distribution is not expected to be the same as the distribution
we refer to as the hitting distribution. There are no predictions
for this distribution defined using the SAW in the full-plane or space, 
and we will not consider it in this paper.

Lawler, Schramm, Werner gave a prediction for the hitting density
of a SAW in two dimensions \cite{lsw_saw}.
Let $D$ be a simply connected domain containing the origin and 
$f$ a conformal map such that $f(0)=0$. We denote the hitting density
with respect to arc length by $\sigma_D(z)$ where $z \in \partial D$. 
Their prediction is that 
\bea
\sigma_D(z) \propto  \sigma_{f(D)}(f(z)) \, |f^\prime(z)|^b
\label{covariance}
\eea
In particular, by taking $f$ to be a conformal map that sends $D$ to 
the unit disc, we see that $\sigma_D(z)$ is proportional to the 
density for harmonic measure on $\partial D$ 
raised to the $b$ power.
Depending on just how one defines the ensemble of SAW's that end 
on the boundary of $D$, there is a correction factor 
that must be included in the above prediction 
\cite{kennedy_lawler}.  It arises from 
lattice effects that, somewhat surprisingly, persist in the scaling 
limit. This is discussed further in section \ref{sect_lat}.

A heuristic derivation of \reff{covariance} can be found in 
\cite{kennedy_3d_prl}. The same argument holds in three 
dimensions with the caveat
that the group of conformal transformations in three dimensions
is quite limited. It is generated by 
translations, rotations, dilations and inversions in spheres.
We will use the conformal transformation 
\beann
\map(x,y,z) ={2(x,y,1-z) \over x^2 + y^2 + (1-z)^2} 
\label{cmap}
\eeann
It maps the point at $\infty$ to the origin, maps the origin to the point
$(0,0,2)$ and maps the unit sphere centered at the origin to the plane 
$z=1$. Note that this is the plane that bisects
the line segment between the two endpoints, so we will often 
refer to it as the bisecting plane.  

The first prediction we test is the hitting density for a SAW in 
the half-space $z<1$, starting at the origin and ending on the 
plane $z=1$. To derive the distribution of the endpoint on the plane
we consider a SAW in the unit sphere centered at the origin which 
starts at the origin and ends on the surface of the sphere. 
The scaling limit of the hitting density should be uniform on 
the sphere. Under the conformal map \reff{cmap}
we have the SAW in the half-space $z<1$ starting at the origin. 
So if we let $(x,y,1)$ be the coordinates of 
a point on the plane $z=1$, then the density with respect to 
Lebesgue measure on the plane is 
\bea
\sigma(x,y) \propto [x^2+y^2 + 1]^{-b} 
\label{hit_plane}
\eea
The constant of proportionality is simply determined by the requirement 
that this is a probability density.
(Details of this derivation can be found in \cite{kennedy_3d_prl}.)
Given the symmetry of the hitting density on the plane 
under rotations about the z-axis, we will just
study the random variable $R=\sqrt{x^2+y^2}$. Its density is proportional 
to $r [r^2+1]^{-b}$, and so the cumulative distribution function (CDF)
of $R$ is $P(R \le r) =  1-(r^2+1)^{1-b}$. The range of $R$ is unbounded, 
so it will be more convenient to display the simulation results
using the angle $\Theta$ given by $\tan \Theta=R$. Its CDF is 
\bea
P(\Theta \le \theta) = 1 - (\cos \theta)^{2(b-1)}
\label{cdf_half}
\eea

The second prediction we test is the hitting density for a SAW 
in a sphere when the starting point is not the origin. 
Without loss of generality we 
can consider the unit sphere centered at the origin and 
take the starting point to be $(0,0,a)$ with $-1 < a < 1$.
We parametrize the sphere with spherical coordinates, and 
let $\rho_a(\theta,\phi)$ be the hitting density with respect to 
Lebesgue measure on the sphere.
Now we use the same conformal map \reff{cmap} as before. 
It still takes the unit sphere to the plane $z=1$, and takes the 
starting point of the SAW at $(0,0,a)$ to $(0,0,2/(1-a))$.
Using the results for the hitting density in a half-space we find 
\bea
\rho_{a}(\theta,\phi) \propto
\left[ 1 + a^2 - 2a \cos(\theta) \right]^{-b} 
\label{hit_sphere}
\eea
(Details of this derivation can be found in \cite{kennedy_3d_prl}.)
If we only consider the angle $\Theta$, the 
density is 
\beann
\rho_{a}(\theta) \propto
\left[ 1 + a^2 - 2a \cos(\theta) \right]^{-b} \, \sin(\theta)
\eeann
Note that this can be explicitly integrated to give the CDF:
\bea
P(\Theta \le \theta) = { (1-a)^{2(1-b)} - (1+a^2-2a \cos(\theta))^{1-b}
\over (1-a)^{2(1-b)} - (1+a)^{2(1-b)}}
\label{cdf_sphere}
\eea

The third prediction we test is for the SAW in the full space starting
at the origin and ending at a fixed point which we take to be $(0,0,2)$. 
The prediction is for the distribution of the point where the walk first
hits the plane $z=1$. 
To derive this distribution we 
consider the scaling limit of the SAW in the full-space from 
the origin to $\infty$. 
Look at the first time it hits the unit sphere. This point will
be uniformly distributed over the sphere. The conformal map
\reff{cmap} transforms this into the
SAW in the full-space from $(0,0,0)$ to $(0,0,2)$ and we are 
looking at the first hit of the plane $z=1$. So its distribution will 
just be the image of the uniform measure on the sphere under $\map$. 
This is given by 
\beann
\sigma_{first}(x,y) =\frac{1}{\pi} [x^2+y^2+1]^{-2} 
\eeann
Let $R=\sqrt{x^2+y^2}$, so $R$ is the distance from this first 
hit of the plane to the center of the plane. 
Then the density of $R$ is proportional to $[r^2+1]^{-2} \, r \, dr$.
So $P(R \le r) = \frac{r^2}{r^2+1}$.
Again, we find it convenient to display our simulation results 
using the angle $\Theta$ given by $\tan \Theta = R$. Its CDF is 
\bea
P(\Theta \le \theta) = \sin^2 \theta
\label{cdf_bisect}
\eea

\section{Equivalence of ensembles}
\label{sect_dilation}

The most efficient algorithm for simulating the SAW is the pivot 
algorithm \cite{madras_slade}.
Clisby's recent implementation of this algorithm
has dramatically improved its efficiency \cite{clisby}.
The pivot algorithm simulates SAW ensembles with 
a fixed number of steps in the walk. The starting point of the walk
is fixed, but the terminal point of the walk is not. In all of the 
ensembles we wish to study the number of steps is not fixed, and the 
terminal point of the walk is either fixed or constrained to lie 
on some boundary. So we cannot directly test the predictions of 
conformal invariance from the previous section with the pivot 
algorithm. Instead we will use the pivot algorithm to study 
ensembles that we believe to have the same scaling limit as the 
ensemble we considered in the previous section, but which are amenable
to simulation by the pivot algorithm. The idea behind these ensembles 
was introduced in \cite{kennedy_dilation}.
Although our focus is on three dimensions, the arguments 
in this section hold in any number of dimensions. 

There are three SAW ensembles used in the paper. The ``half-space
ensemble'' consists of all finite length SAW's that start at the origin,
and then stay in the half-space $z<1$ until they end 
somewhere on the plane $z=1$. The ``spherical ensemble'' consists
of all finite length SAW's that start at the origin and then 
stay inside the sphere of radius $1$ centered at $(0,0,-a)$ 
until they end somewhere on the surface of this sphere.
The ``point to point ensemble'' consists of all finite length SAW's 
in the full-space that start at the origin and end at a fixed point $q$. 
Note that all three of these ensembles 
share two features: there is no constraint on the number of steps
in the walk, but there is a constraint on where the walk ends. 

The pivot algorithm simulates the SAW ensemble which contains all 
SAW's starting at the origin with a fixed number $N$ of steps and no 
constraint on where they end. 
We refer to this as the ``fixed-length ensemble'',  
and denote expectations with respect to the uniform probability measure
on these walks by $E_N$. The pivot algorithm can also be used to simulate
the ensemble of SAW's with $N$ steps which start at the origin and then 
stay in the half-space $z>0$. 
We refer to this as the ``half-space fixed-length ensemble'', and 
denote expectations with respect to the uniform probability measure
on it by $E_N^+$. 
In this section we show 
how to express the three ensembles used in this paper in terms
of the fixed-length ensembles. 
In all three cases the general strategy is the same. We define 
a ``super-ensemble'' in which the number of steps in the SAW
is not fixed and there are no constraints on where the SAW ends. 
We add a constraint to the super-ensemble that says 
that it is possible to transform the walk 
into a walk which appears in the desired ensemble 
by a Euclidean transformation. We then decompose the sum in 
the super-ensemble in two different ways. One decomposition shows
that its scaling limit is the same as the scaling limit of the desired
ensemble. The other decomposition shows that its scaling limit 
can be expressed in terms of the $N \ra \infty$ limit of the 
fixed-length ensemble.

\subsection{Half-space ensemble}
\label{sect_half_space}

We first consider the half-space ensemble which consists of 
all SAW's on a lattice with spacing
$\delta$ that start at the origin and then stay in the half-space
$\{(x,y,z): z<1\}$ until they end on the boundary of this half-space, 
i.e., the plane $z=1$. 
A walk $\omega$ is given weight $\mu^{-|\omega|}$. 
We denote expectations in this half-space ensemble with lattice
spacing $\delta$ by $E^{half}_\delta$. 
We want to express this half-space ensemble 
in terms of the half-space fixed-length ensemble which 
consists of all SAW's with $N$ steps in the half-space $z>0$
that start at the origin. All the walks are given equal probability
in the half-space fixed-length ensemble.
We denote expectations in the half-space fixed-length ensemble
by $E^+_\fixn$.
Given a walk from the half-space fixed-length ensemble we can 
reflect it in the plane $z=0$, then translate it and dilate it to 
produce a walk between the origin and the plane $z=1$ which
stays in the half-space $z<1$, i.e., a walk in the half-space ensemble.
If $\omega$ is the walk from the half-space fixed-length ensemble, 
we let $\omega(|\omega|)=(x_0,y_0,z_0)$ and then let 
\beann
\phi_\omega(x,y,z)=({x-x_0 \over z_0},{y-y_0 \over z_0},{z_0-z \over z_0})
\eeann
Then $\phi_\omega$ is the Euclidean transformation that takes the endpoint
of $\omega$ to the origin and takes the origin to some point on the plane
$z=1$. So by reversing its direction 
we can think of $\phi_\omega(\omega)$ as a walk 
in the half-space ensemble.
Our conjectured relationship between the half-space ensemble and the 
half-space fixed-length ensemble is as follows. 

\smallskip

\no {\bf Equivalence of ensembles for the half-space:} 
Let $\psi(\gamma)$ be a random variable on the space of 
simple curves $\gamma$ in the half-space $z<1$ which start at the origin and 
end somewhere on the plane $z=1$. Then 
\bea
\lim_{\delta \ra 0} E_\delta^{half}(\psi) 
= \lim_{\fixn \ra \infty} 
\frac{E^+_\fixn [ z(\walk)^p \psi(\phi_\walk(\walk))]}{E^+_\fixn [ z(\walk)^p]}
\label{equiv_half}
\eea
where $p = \frac{\rho-\gamma}{\nu}$.
The function $z(\omega)$ is the $z$-coordinate
of $\omega(N)$, the endpoint of the walk. 
In words, we can simulate SAW's from the half-space ensemble of walks 
that start at the origin and stay in the half-space $z<1$ until they
end on the plane $z=1$ as follows. We generate walks $\walk$ from 
the fixed-length ensemble, apply a Euclidean transformation so 
that the reversal of $\walk$ goes from the origin to the plane $z=1$ 
and weight the walk by $z(\walk)^p$.

\smallskip

To derive this relationship we use the super-ensemble which consists of
all finite SAW's in the half-space $z>0$ that start at the origin. 
A walk $\omega$ is weighted by $\mu^{-|\omega|}$. 
To ensure that the sum of the weights is finite we introduce a cutoff. 
For a walk $\omega$ we let $z(\omega)$ be the $z$-coordinate
of $\omega(|\omega|)$, the endpoint of the walk. 
Fix $0 < z_1 < z_2$. 
Our cutoff is the condition $z_1 \le z(\omega) \le z_2$. 
Define
\bea
\Z_\delta(\psi)= \sum_{\omega:0 \ra \half} \mu^{-|\omega|} 
1(z_1 \le z(\omega) \le z_2) \,
\psi(\phi_\omega(\omega))
\eea
where $\half$ stands for the half-space $z>0$ and 
the notation $\omega:0 \ra \half$ means that the sum 
is over all finite SAW's which start at the origin and then stay in the 
half-space $\half$. 

We now decompose the sum in $\Z_\delta(\psi)$ in two different ways.  
The first decomposition is based on where the walk ends. 
\bea
\Z_\delta(\psi)= \sum_{z_1 \le z \le z_2} 
\quad \sum_{\omega:0 \ra \half, z(\omega)=z} 
\mu^{-|\omega|} \, \psi(\phi_\omega(\omega))
\eea
The sum over $z$ is over multiples of $\delta$ that are in the allowed range.
Define 
\beann
Z_{\delta,z}= \sum_{\omega:0 \ra \half, z(\omega)=z} \mu^{-|\omega|} \, 
\eeann
Consider
\beann
{1 \over Z_{\delta,z}} \sum_{\omega:0 \ra \half, z(\omega)=z} 
\mu^{-|\omega|} \, \psi(\phi_\omega(\omega)) 
\eeann
Thanks to the constraint $z(\omega)=z$, the transformations $\phi_{\omega}$ 
in this sum all have the same dilation factor, namely $1/z$. 
So the walks $\phi_\omega(\omega)$ all live on a lattice with spacing 
$\delta/z$. They all start at the origin and stay in the half-space 
$z<1$ until they end on the plane $z=1$.  
So the above is exactly
equal to the expectation of $\psi$ in the half-space ensemble with 
lattice spacing $\delta/z$, i.e.,  $E^{half}_{\delta/z}[\psi]$. 
For all $z$, as $\delta \ra 0$, this has the same limit 
as $E^{half}_\delta[\psi]$. 
Noting that 
\beann
\sum_{z_1 \le z \le z_2} Z_{\delta,z} = \Z_\delta(1)
\eeann
we conclude that 
\bea
\lim_{\delta \ra 0} {\Z_\delta(\psi) \over \Z_\delta(1)} 
=\lim_{\delta \ra 0} E^{half}_\delta [\psi]
\label{halfa}
\eea

For the second decomposition of $\Z_\delta(\psi)$, 
we decompose it based on the number of steps in $\omega$. 
Let $b_n$ be the number of SAW's in the half-space starting at $0$ 
with $n$ steps. We rewrite $\Z_\delta(\psi)$ as 
\beann
\Z_\delta(\psi) &=& \sum_{\varn=1}^\infty \mu^{-n} b_n \, {1 \over b_n}
\sum_{\omega: 0 \ra \half,|\omega|=\varn} \, 
1(z_1 \le z(\omega) \le z_2) 
\, \psi(\phi_\omega(\omega)) \\
&=& \sum_{\varn=1}^\infty \mu^{-\varn} b_\varn \, 
E^+_\varn[1(z_1 \le z(\omega) \le z_2) \, \psi(\phi_\omega(\omega))]
\eeann
Since $z(\omega) \ge z_1$, $\omega$ must
have at least $z_1/\delta$ steps. So as the lattice spacing goes 
to zero, the first $n$ for which the summand is nonzero
goes to infinity. 
Since $b_\varn$ is asymptotic to $\mu^{\varn} \varn^{\gamma-1-\rho}$, 
we replace $b_\varn \mu^{-\varn}$ by $\varn^{\gamma-1-\rho}$.
So we have 
\beann
\Z_\delta(\psi) \approx \sum_{\varn=1}^\infty \varn^{\gamma-1-\rho} \, 
E^+_\varn[1(z_1 \le z(\omega) \le z_2) \, \psi(\phi_\omega(\omega)) ]
\eeann

Since  the sum on $\varn$ is only over large values, 
we can replace the expectations $E^+_\varn$ over $\omega$ 
with a single expectation $E^+_\fixn$ over $\walk$ where $\fixn$ is large. 
We replace $\omega \varn^{-\nu}$ with $\walk \fixn^{-\nu}$.
So $z(\omega)$ becomes 
$z(\varn^\nu \fixn^{-\nu} \walk) = \varn^\nu \fixn^{-\nu} z(\walk)$.
And we replace $\psi(\phi_\omega(\omega))$ by 
$\psi(\phi_\walk(\walk))$.
We now have
\beann
\Z_\delta(\psi) \approx \sum_{\varn=1}^\infty \varn^{\gamma-1-\rho} 
\, E^+_\fixn [ 
\,  1(z_1 \le \varn^\nu \fixn^{-\nu} z(\walk) \le z_2) 
\, \psi(\phi_\walk(\walk))]
\eeann

The $\varn$ dependent part of the preceding sum is 
\beann
\sum_{\varn=1}^\infty \varn^{\gamma-1-\rho} \, \,
 1(z_1 \le \varn^\nu \fixn^{-\nu} z(\walk) \le z_2)
\eeann
If we multiply this by $\fixn^{\rho-\gamma}$ 
it becomes a Riemann sum approximation
to an integral, and we see that as $\fixn \ra \infty$, 
\beann
&& \fixn^{\rho-\gamma} \sum_{\varn=1}^\infty \varn^{\gamma-1-\rho} \, \,
 1(z_1 \le \varn^\nu \fixn^{-\nu} z(\walk) \le z_2)
= 
\frac{1}{\fixn} \sum_{\varn=1}^\infty 
\left(\frac{\varn}{\fixn}\right)^{\gamma-1-\rho} \, \,
 1(z_1 \le \left( \frac{\varn}{\fixn} \right)^\nu z(\walk) \le z_2) \\
&& \ra \int_0^\infty x^{\gamma-1-\rho} \, 
1(z_1 \le x^\nu z(\walk) \le z_2)
\, dx = c \, [z(\walk)]^{(\rho-\gamma)/\nu}
\eeann
where the constant $c$ depends on $z_1,z_2,\rho,\gamma,\nu$.
Thus
\bea
\lim_{\delta \ra 0} \frac{\Z_\delta(\psi)}{\Z_\delta(1)}
= \lim_{\fixn \ra \infty} 
\frac{E^+_\fixn [ z(\walk)^p \psi(\phi_\walk(\walk))]}{E^+_\fixn [ z(\walk)^p]}
\label{halfb}
\eea
with $p = \frac{\rho-\gamma}{\nu}$.
Combining eqs. \reff{halfa} and \reff{halfb} we obtain eq. \reff{equiv_half}.

\subsection{Spherical ensemble}
\label{sect_sphere}

The second ensemble in the paper consists of 
all finite length SAW's that start at $(0,0,a)$ and then 
stay inside the sphere of radius $1$ centered at the origin until 
they end somewhere on the surface of this sphere.
(The parameter $a$ satisfies $-1<a<1$.)
By a simple translation this becomes the ensemble of all walks that
start at the origin and then stay inside the sphere of radius $1$ 
centered at $(0,0,-a)$ until they end somewhere on the surface of this
sphere. A walk $\omega$ is given weight $\mu^{-|\omega|}$. 
We will refer to this ensemble of walks starting at the origin as 
the spherical ensemble, and denote expectations in 
this ensemble by $E^{sphere}_\delta$.
We want to express this spherical ensemble in terms of the fixed-length
ensemble which consists of all $N$ step walks in the full space which
start at the origin. The walks are given equal probability in this
ensemble, and we denote expectations by $E_\fixn$.
Given a walk $\omega$ from the fixed-length ensemble,
there is a dilation factor $R(\omega)$ such 
that the dilated walk $\omega/R(\omega)$ will end on the 
surface of the sphere centered at $(0,0,-a)$ with radius $1$. 
In general this dilated walk will not stay inside the sphere, 
so we must condition on the event that it does. 
We let $1_S(\omega/R(\omega))$ denote the indicator function for this event. 
Our conjectured relationship between the spherical ensemble and the 
fixed-length ensemble is as follows. 

\smallskip

\no {\bf Equivalence of ensembles for the sphere:} 
Let $\psi(\gamma)$ be a random variable on the space of 
simple curves $\gamma$ from the origin to the boundary of the unit sphere 
centered at $(0,0,-a)$. Then 
\bea
\lim_{\delta \ra 0} E_\delta^{sphere}[\psi]
= \lim_{\fixn \ra \infty} 
\frac{E_\fixn [ R(\walk)^p \, 1_S(\frac{\walk}{R(\walk)}) \, W(\walk)
\psi(\frac{\walk}{R(\walk)})]}
{E_\fixn [ R(\walk)^p 1_S(\frac{\walk}{R(\walk)}) \, W(\walk)] }
\label{equiv_sphere}
\eea 
where $p = \frac{\rho-\gamma}{\nu}$.
The function $W(\walk)$ depends only on the endpoint of $\walk$ and 
will be defined later.
So we can simulate SAW's from the spherical ensemble of walks 
that start at the origin and stay in the sphere of radius $1$ centered 
at $(0,0,-a)$ until they end on its boundary 
as follows. We generate walks $\walk$ from the fixed-length ensemble. 
We rescale the walks so that they end on the given sphere and then 
condition on the event that the walk stays inside the sphere. 
These walks are then weighted by $R(\walk)^p W(\walk)$.

\smallskip

To derive this relationship  we introduce the following  super ensemble.
We start with all finite SAW's that start at the origin.
A walk $\omega$ is weighted by $\mu^{-|\omega|}$. 
Fix $0 < R_1 < R_2$. 
Define
\bea
\Z_\delta(\psi)= \sum_{\omega:0 \ra } \mu^{-|\omega|} 
1(R_1 \le R(\omega) \le R_2) \, 1_S(\frac{\omega}{R(\omega)}) \,
W(\omega) \, \psi(\frac{\omega}{R(\omega)}) 
\label{R_psi}
\eea
where the sum is over all finite SAW's starting at the origin.
$1_S(\frac{\omega}{R(\omega)})$ is the indicator function that the  dilated 
walk stays in the sphere. We have introduced a cutoff by 
requiring $R(\omega)$ to lie between $R_1$ and $R_2$.
Let $||\omega(|\omega|)||$ be the distance to 
the endpoint of $\omega$, and let $\theta$ be the angle between the $z$-axis
and the line from the origin to the endpoint of $\omega$.
Some calculation shows
\beann
R(\omega)= ||\omega(|\omega|)|| 
\frac{a \cos(\theta) + \sqrt{1-a^2\sin^2(\theta)}}{1-a^2}  \\
\eeann

We now decompose the sum in $\Z_\delta(\psi)$ in two different ways.  
The first decomposition is based on $R(\omega)$. We fix an $R$ and 
an infinitesimal $dR$ and consider the 
following ``slice'' of $Z_\delta(\psi)$. 
\bea
\sum_{\omega:0 \ra } \mu^{-|\omega|} 
1(R \le R(\omega) \le R + dR) \, 
1_S(\frac{\omega}{R(\omega)}) \, W(\omega) \, \psi(\frac{\omega}{R(\omega)}) 
\eea
The definition of $R(\omega)$ implies that $\omega$ ends on the surface
of the sphere of radius $R(\omega)$ centered at $(0,0,-aR(\omega))$. 
So the sum above contains all walks that start at the origin and end in 
a thin shell about the surface of 
the sphere of radius $R$ centered at $(0,0,-aR)$. 
If the thickness of this shell were uniform and we did not include the 
factor of $W(\omega)$, then in the scaling limit the above sum with 
the appropriate normalization  would converge to the expectation of $\psi$ in 
the spherical ensemble. However, the thickness of the shell is not
uniform. We correct for this by taking $W(\omega)$ to be proportional to 
the reciprocal of the shell thickness. So $W(\omega)$ only depends
on $\omega$ through its endpoint; in fact it is only a function of 
$\theta$. We can think of $Z_\delta(\psi)$ as a sum over many 
shells, and so we conclude that 
with this definition of $W(\omega)$, 
\bea
\lim_{\delta \ra 0} \frac{Z_\delta(\psi)}{Z_\delta(1)} =
\lim_{\delta \ra 0} E_\delta^{sphere}[\psi]
\label{spherea}
\eea

For the second decomposition of $\Z_\delta(\psi)$, 
we decompose it based on the number of steps in $\omega$. 
Let $c_\varn$ be the number of SAW's starting at $0$ 
with $\varn$ steps. We rewrite $\Z_\delta(\psi)$ as 
\beann
\Z_\delta(\psi) 
&=& \sum_{\varn=1}^\infty \mu^{-\varn} c_\varn \, {1 \over c_\varn}
\sum_{\omega: 0 \ra ,|\omega|=\varn} \, 
1(R_1 \le R(\omega) \le R_2) \, 
1_S(\frac{\omega}{R(\omega)}) \, W(\omega) \,
\psi(\frac{\omega}{R(\omega)}) \\
&=& \sum_{\varn=1}^\infty \mu^{-\varn} c_\varn \, 
E_\varn[1(R_1 \le R(\omega) \le R_2) \, 
1_S(\frac{\omega}{R(\omega)}) \, W(\omega) \psi(\frac{\omega}{R(\omega)})]
\eeann
As the lattice spacing goes 
to zero, the first $\varn$ for which the summand in the sum on $\varn$ 
is nonzero goes to infinity. 
Since $c_\varn$ is asymptotic to $\mu^{\varn} \varn^{\gamma-1}$, 
we replace $\mu^{-\varn} c_\varn$ by $\varn^{\gamma-1}$.
As in our previous derivation we replace the expectations 
$E_\varn$ by a single expectation $E_\fixn$ 
by replacing $\varn^{-\nu} \omega$ with $\fixn^{-\nu} \walk$ where 
$\walk$ comes from $E_\fixn$.
So $R(\omega)$ becomes 
$R(\varn^\nu \fixn^{-\nu} \walk) = \varn^\nu \fixn^{-\nu} R(\walk)$.
We can replace $\psi(\frac{\omega}{R(\omega)})$ by 
$\psi(\frac{\walk}{R(\walk)})$. 
The indicator function $1_S(\frac{\omega}{R(\omega)})$
is more subtle. The probability that an $\varn$-step SAW stays 
on one side of a half-plane is conjectured to go to zero as $\varn^{-\rho}$ 
as $\varn \ra \infty$. So we expect
that the probability that $1_S(\frac{\omega}{R(\omega)})=1$
also goes to zero as $\varn^{-\rho}$. 
So we approximate $\varn^\rho 1_S(\frac{\omega}{R(\omega)})$ 
by $\fixn^\rho 1_S(\frac{\walk}{R(\walk)})$ 
i.e., we replace $1_S(\frac{\omega}{R(\omega)})$ 
by $\varn^{-\rho} \fixn^\rho 1_S(\frac{\walk}{R(\walk)})$.
We now have
\beann
\Z_\delta(\psi) \approx \fixn^\rho 
\sum_{\varn=1}^\infty \varn^{\gamma-1-\rho} \, 
E_\fixn [\,  1(R_1 \le \varn^\nu \fixn^{-\nu} R(\walk) \le R_2) 
1_S(\frac{\walk}{R(\walk)}) \, W(\walk) \, \psi(\frac{\walk}{R(\walk)})]
\eeann

The $\varn$ dependent part of this is 
\beann
\sum_{\varn=1}^\infty \varn^{\gamma-1-\rho} \, \,
 1(R_1 \le \varn^\nu \fixn^{-\nu} R(\walk) \le R_2)
\eeann
If we multiply this by $\fixn^{\rho-\gamma}$ it becomes a Riemann sum 
approximation to an integral, and so as $\fixn \ra \infty$, 
\beann
\fixn^{\rho-\gamma}
\sum_{\varn=1}^\infty \varn^{\gamma-1-\rho} \, \,
 1(R_1 \le \varn^\nu \fixn^{-\nu} R(\walk) \le R_2)
\ra \int_0^\infty x^{\gamma-1-\rho} \, 
 1(R_1 \le R(\walk) x^{\nu} \le R_2) \, dx 
= c R(\walk)^{(\rho-\gamma)/\nu}
\eeann
Thus
\bea
\lim_{\delta \ra 0} \frac{\Z_\delta(\psi)}{\Z_\delta(1)}
= \lim_{\fixn \ra \infty} 
\frac{E_\fixn [ R(\walk)^p \, 1_S(\frac{\walk}{R(\walk)}) \, W(\omega)
\psi(\frac{\walk}{R(\walk)})]}
{E_\fixn [ R(\walk)^p 1_S(\frac{\walk}{R(\walk)}) \, W(\omega)] }
\label{sphereb}
\eea
where $p = \frac{\rho-\gamma}{\nu}$.
Combining eqs. \reff{spherea} and \reff{sphereb}
we obtain eq. \reff{equiv_sphere}.

\subsection{Point to point ensemble}
\label{sect_pt_to_pt}

Finally, we consider the point to point ensemble of SAW's.
We take one point to be the origin and label the other point as $q$. 
The ensemble consists of all SAW's on the lattice 
which start at $0$ and end at $q$. (We interpret ending at $q$ 
as meaning the walk ends at the nearest point in the lattice to $q$.) 
All finite length SAW's between the points are allowed, and the probability 
of a walk $\omega$ is proportional to $\mu^{-|\omega|}$
where $|\omega|$ is the number of steps in the walk. 
We let $E_\delta^{0 \ra q}$ denote expectation with respect to this ensemble. 
As before $E_\fixn$ denotes expectation in the fixed-length ensemble
of walks in the full plane starting at the origin. 
Given a point $z$, in two dimensions there is a unique Euclidean symmetry 
that maps it to $q$ and fixes the origin. In three and higher dimensions,
there are many Euclidean symmetries that do so. 
For each $z$ we pick one and denote it by $\phi_z$. 
Given a finite walk $\omega$ starting at  $0$, 
$\phi_{\omega(|\omega|)}$ transforms $\omega$ into a walk from the origin 
to $q$. To simplify the notation we denote $\phi_{\omega(|\omega|)}$ just by 
$\phi_\omega$, but we emphasize that it only depends on the endpoint of 
$\omega$.

\smallskip

\no {\bf Equivalence of ensembles for the full space:} 
Let $\psi(\gamma)$  be an observable on 
the space of simple curves $\gamma$ from $0$ to $q$. 
Then 
\bea
\lim_{\delta \ra 0} E_\delta^{0 \ra q}(\psi) 
= \lim_{\fixn \ra \infty} {E_\fixn [ ||\walk(\fixn)||^{-\gamma/\nu} 
\psi(\phi_\walk(\walk))] \over E_\fixn [ ||\walk(\fixn)||^{-\gamma/\nu} ]}
\label{equiv_full}
\eea
In words, we can simulate SAW's from the ensemble of walks 
between $0$ and $q$ by generating walks $\walk$ from 
the fixed-length ensemble, applying a Euclidean transformation so 
that the walk goes between $0$ and $q$ and weighting the walk by 
$||\walk(\fixn)||^{-\gamma/\nu}$.

\smallskip

To derive this relationship we use the super-ensemble 
consisting of all finite length 
SAW's which start at the origin. The total mass of this ensemble is 
infinite, so we introduce a cutoff.
Fix $0 < r_1 < r_2$. The cutoff is that $r_1 \le ||\omega(|\omega|)|| \le r_2$.
We define
\beann
\Z_\delta(\psi) = \sum_{\omega:0 \ra} \mu^{-|\omega|} 
\, 1(r_1 \le ||\omega(|\omega|)|| \le r_2) \, \psi(\phi_\omega(\omega))
\eeann
where the sum is over all finite walks 
starting at $0$ on a lattice with spacing $\delta$.
$\omega(|\omega|)$ is the endpoint of the walk, and $||\omega(|\omega|)||$
denotes the usual Euclidean distance of this point from the origin. 

First we decompose the sum over walks in $\Z_\delta(\psi)$ based on where the 
walk ends:
\beann
\Z_\delta(\psi) = \sum_z \sum_{\omega:0 \ra z} \mu^{-|\omega|} \, 
1(r_1 \le ||z|| \le r_2) \, \psi(\phi_\omega(\omega))
\eeann
Define
\beann
Z_{\delta,z}= \sum_{\omega:0 \ra z} \mu^{-|\omega|} 
\eeann 
So $Z_{\delta,z}$ is the partition function corresponding to $E_\delta^{0 \ra z}$.
Then 
\beann
\Z_\delta(\psi)  &=& \sum_{z: r_1 \le ||z|| \le r_2} \,  Z_{\delta,z} \, 
E_\delta^{0 \ra z} [\psi(\phi_z(\omega))]
\eeann
We expect the scaling limit to have euclidean invariance, and 
so in the scaling limit \\
$E_\delta^{0 \ra z} [\psi(\phi_z(\omega))]$  and 
$E_\delta^{0 \ra q} [\psi(\omega)]$ should converge to the same limit 
for all $z$. 
The sum of $Z_{\delta,z}$ over $z$ satisfying $r_1 \le ||z|| \le r_2$
is just $\Z_\delta(1)$.
So 
\bea
\lim_{\delta \ra 0} \frac{\Z_\delta(\psi)}{\Z_\delta(1)} =
\lim_{\delta \ra 0} E_\delta^{0 \ra q} [\psi(\omega)]
\label{fulla}
\eea

Our second decomposition of $\Z_\delta(\psi)$ is based on the number of 
steps in the walk. Letting $c_\varn$ denote the number of SAW's 
starting at the origin with $\varn$ steps, we have
\beann
\Z_\delta(\psi) &=& 
\sum_{\varn=1}^\infty \mu^{-\varn} c_\varn \, {1 \over c_\varn}
\sum_{\omega:0 \ra, |\omega|=\varn } \, 1(r_1 \le ||\omega(\varn)|| \le r_2)
\, \psi(\phi_\omega(\omega)) \\
&=&
\sum_{\varn=1}^\infty \mu^{-\varn} c_\varn  \,
E_\varn [1(r_1 \le ||\omega(\varn)|| \le r_2) \, \psi(\phi_\omega(\omega))]
\eeann
The constraint that the endpoint of the walk is at a distance at least
$r_1$ implies that as the lattice spacing goes to zero, the 
terms in the above sum are non-zero only for larger and larger
$\varn$. Since $c_\varn$ is expected to be asymptotic to 
$\mu^\varn \varn^{\gamma-1}$, we replace $\mu^{-\varn} c_\varn$ by 
$\varn^{\gamma-1}$. 
For large $\varn$ we can approximate the expectations $E_\varn$ by
a single expectation $E_\fixn$ where $\fixn$ is large if we rescale the walks
suitably. 
More precisely, if we sample $\omega$ from $E_\varn$ and $\walk$ from 
$E_\fixn$, then
$\omega \varn^{-\nu}$ and $\walk \fixn^{-\nu}$ should have approximately 
the same distribution when $\varn$ and $\fixn$ are large. 
Since 
$\phi_{\omega \varn^{-\nu}}(\omega \varn^{-\nu})=\phi_\omega(\omega)$,
$\psi(\phi_\omega(\omega))$
just becomes $\psi(\phi_\walk(\walk))$. So 
\beann
\Z_\delta(\psi) = \sum_{\varn=1}^\infty \mu^{-\varn} \varn^{\gamma-1} 
E_\fixn\left[ 1(r_1 \le ||\walk(\fixn)|| \varn^{\nu} \fixn^{-\nu} \le r_2) 
\, \psi(\phi_\walk(\walk)) \right]
\eeann
The terms in the above that depend on $\varn$ are 
\beann
\sum_{\varn=1}^\infty \varn^{\gamma-1} \, 1(r_1 \le ||\walk(\fixn)|| 
\varn^{\nu} \fixn^{-\nu} \le r_2) 
\eeann
If we multiply this by $\fixn^{-\gamma}$ then as $N \ra \infty$ 
\beann
\fixn^{-\gamma} \sum_{\varn=1}^\infty 
\varn^{\gamma-1} \, 1(r_1 \le ||\walk(\fixn)|| 
\varn^{\nu} \fixn^{-\nu} \le r_2) 
\ra \int_0^\infty x^{\gamma-1} \, 
 1(r_1 \le ||\walk(\fixn)|| x^{\nu} \le r_2) \, dx 
= c ||\walk(\fixn)||^{-\gamma/\nu}
\eeann
So 
\bea
\lim_{\delta \ra 0} \frac{\Z_\delta(\psi)}{\Z_\delta(1)}
= \lim_{\fixn \ra \infty} {E_\fixn [ ||\walk(\fixn)||^{-\gamma/\nu} 
\psi(\phi_\walk(\walk))] \over E_\fixn [ ||\walk(\fixn)||^{-\gamma/\nu} ]}
\label{fullb}
\eea
Eq. \reff{equiv_full} follows from eqs. \reff{fulla} and \reff{fullb}.

\section{Lattice effect function}
\label{sect_lat}

The prediction \reff{hit_sphere} for the hitting density for the 
sphere is not exactly correct. There is a lattice effect that persists
in the scaling limit that must be taken into account. 
This effect is not present in the other two simulations since the 
surface involved is a plane. 
The exact form of this lattice effect depends on just how one defines
the ensemble of SAW's ending on the boundary of the sphere.
In  \cite{kennedy_lawler} it was conjectured that in two dimensions the 
local lattice effect at a point on the boundary only depends on the 
angle of the tangent to the boundary with respect to the lattice. 
We expect the analogous result holds in three dimensions, i.e., the 
local lattice effect only depends on the orientation of the tangent plane
to the surface with respect to the lattice. 
In the context of our prediction \reff{hit_sphere} for the sphere, this
means it only depends on the spherical angles $\theta$ and $\phi$
of the endpoint of the walk. 

Explicit conjectures for these local lattice effects were given in 
\cite{kennedy_lawler} for two particular definitions of the ensemble
of SAW's ending on the boundary of a domain and for a third definition
in \cite{kennedy_dilation}. The ensemble we consider here is completely 
analogous to the two dimensional ensemble considered in 
\cite{kennedy_dilation}. 
Requiring the SAW to stay inside the dilated sphere and end on its boundary
has both a macroscopic and microscopic effect.
The prediction \reff{hit_sphere} comes from the macroscopic effect. 
Near the endpoint of the walk the microscopic effect is that the SAW 
must stay on one side of the tangent plane to the sphere at the endpoint. 
This will produce a factor $l(\theta,\phi)$ that depends on the angle of 
the tangent line with respect to the lattice. 

Let $P$ be the plane through the origin
which is parallel to the plane that is tangent to the sphere at the 
point with angles $\theta,\phi$.
We consider SAW's with $N$ steps starting at the origin.
Let $c_N$ be the number of such walks, and let
$b_N(\theta,\phi)$ be the number of such walks 
that stay on one side of the plane.
So $b_N(\theta,\phi)/c_N$ is the probability that an 
$N$ step SAW stays on one side of the plane. We expect 
that this probability goes to zero as $N^{-\rho}$ as $N \ra \infty$, 
and we conjecture that the lattice effect is given by the function
\bea
l(\theta,\phi)= \lim_{N \ra \infty} {b_N(\theta,\phi) \over c_N} N^{\rho}
\eea
So if we use the weight $W(\omega)$ in our ensemble as we did in \reff{R_psi}, 
then the hitting density will be 
$\rho_{a}(\theta,\phi) l(\theta(\omega),\phi(\omega))$, 
rather than $\rho_{a}(\theta,\phi)$ as it should be.
Here $\theta(\omega),\phi(\omega)$ are the spherical angles of the endpoint 
of the walk $\omega$.
To remove this lattice effect from our ensemble we replace
the weight $W(\omega)$ by 
\bea
\hat{W}(\omega)={W(\omega) \over l(\theta(\omega),\phi(\omega))}
\label{weight_hat}
\eea

We do not have an explicit conjecture for $l(\theta,\phi)$. 
We must estimate it by a separate Monte Carlo simulation.
Since $\rho_{a}(\theta,\phi)$ only depends on $\theta$, we 
do not need to compute the full function $l(\theta,\phi)$. 
We only need to compute  
\beann
\hat{l}(\theta)=\int_0^{2 \pi} \, l(\theta,\phi) \, d \phi
\eeann
The integral over $\phi$ can be done as part of the simulation. 
We generate a large number of SAW's with $N$ steps starting at the origin. 
For each SAW we randomly pick a $\phi$ uniformly from $[0,2 \pi]$. 
For a fixed SAW and $\phi$ there will be an interval of $\theta$, 
possibly empty, for which the SAW stays on one side of the plane 
through the origin that is parallel to the tangent plane to the sphere
at the point with spherical coordinates $\theta,\phi$. The average of 
the indicator function of this interval over the SAW samples is 
then an approximation for $\hat{l}(\theta)$. 

\begin{figure}[tbh]
\includegraphics{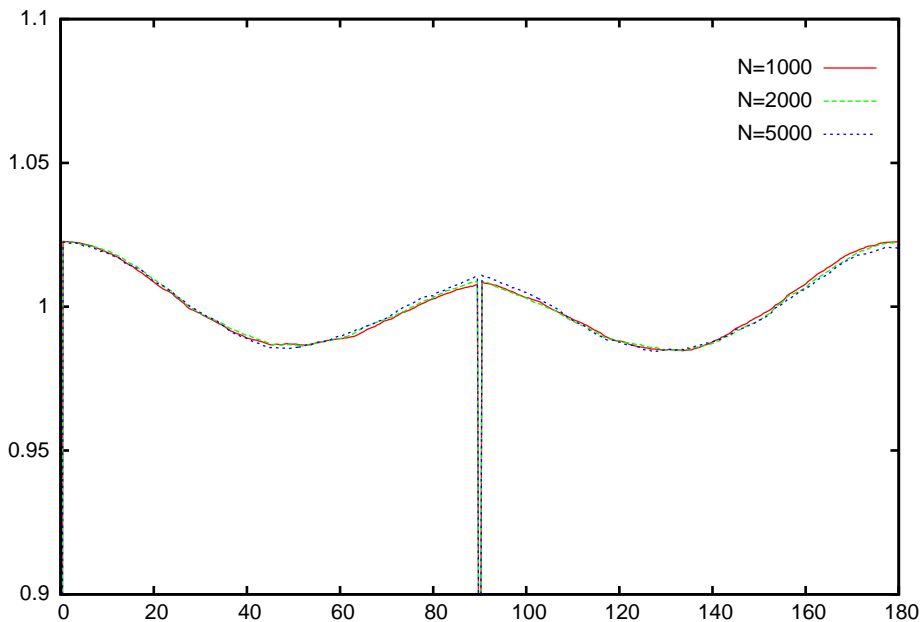}
\caption{\leftskip=25 pt \rightskip= 25 pt 
The lattice effect function using $80$ million samples.
Three curves are shown for $N=1000, 2000, 5000$.
Each curve is rescaled so it has average $1$.
}
\label{fig_lat}
\end{figure}

We carried out this simulation for $N=10,20,50,100,200,500,1000,2000,5000$ 
with 80 million samples for each $N$. 
A sample is a computation of range of $\theta$, which is often empty.  
For the smaller values of $N$ the result depends significantly 
on $N$. In figure \ref{fig_lat} we plot our approximations for 
$\hat{l}(\theta)$ that result from the simulations with $N=1000, 2000$
and $5000$. Each curve has been  rescaled so its average is $1$. 
The figure shows that for these three values of $N$ the approximations
are very close to each other. 
We use the data for $N=1000$ 
when we study the simulations for the spherical ensemble. 

\section{Simulation tests of the predictions}

We use the pivot algorithm to simulate the ensemble of SAW's with 
a fixed number of steps. We use Clisby's implementation of this algorithm
which has dramatically increased its speed \cite{clisby}.
For all our simulations we plot the cumulative distribution
function (CDF) rather than the density. 
Finding the density from a simulation requires taking a numerical
derivative and so adds further uncertainty. 
As can be seen in figure 2 in 
\cite{kennedy_3d_prl}, if we plot the predicted CDF and the CDF 
found in the simulation, the difference between them is too small to 
be seen in such a plot.
So in this paper we only plot the differences - the simulation CDF minus
the predicted CDF.

\begin{figure*}
\includegraphics{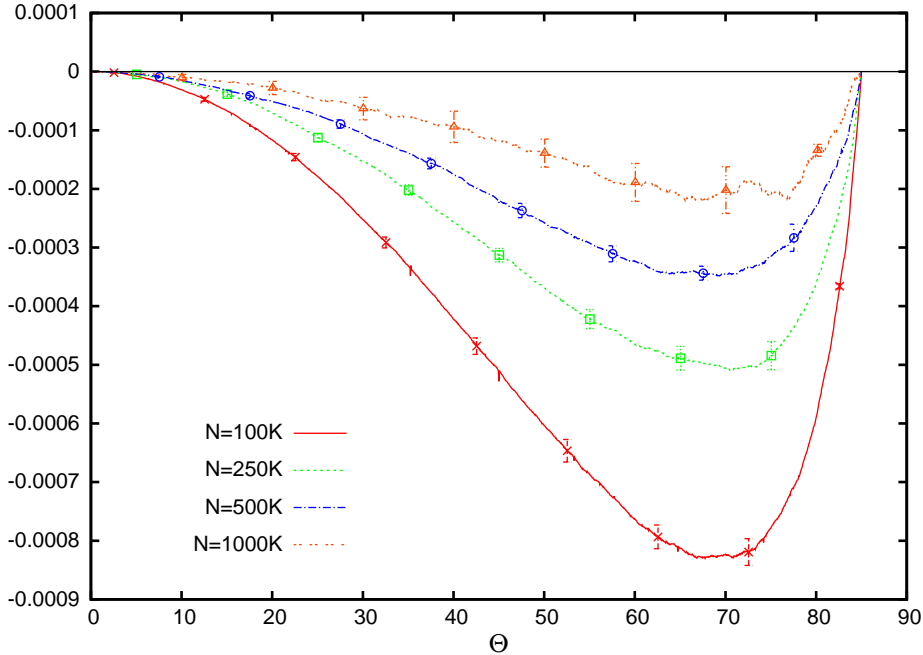}
\caption{
The curves are the difference between the simulation CDF's for 
the hitting density for the half-space and the predicted CDF given 
by eq. \reff{cdf_half}. 
Four differences are shown corresponding to $N=100K, 250K, 500K, 1000K$. 
}
\label{fig_half}
\end{figure*}

The first prediction we test is 
the hitting density for the SAW in the half-space $z<1$ 
starting at the origin and ending on the plane $z=1$.
As discussed in section \ref{sect_half_space} we can study 
this ensemble by using the pivot algorithm to generate 
samples of the half-space fixed-length ensemble.
This ensemble consists of all SAW's with $N$ steps which start 
at the origin and stay in the half-space $z>0$.
For any such SAW we can dilate it and translate it to produce a SAW 
in the half-space $z<1$ that goes
between the origin and the plane $z=1$. 
As discussed in section \ref{sect_half_space}, if we 
weight the SAW $\omega$  by $z(\omega)^{(\rho-\gamma)/\nu}$
then we expect the scaling limit to be the same as the 
scaling limit of the SAW in the half-space $z<1$ 
which start at the origin and end on the plane $z=1$.
Here $z(\omega)$ is the $z$-component of the endpoint of the walk. 
When the SAW $\omega$ ends close to the plane $z=0$, 
$z(\omega)$ will be relatively small. Such SAW's are improbable, 
but the weighting factor is large and so the 
statistical errors in the simulation are increased. 
The troublesome SAW's typically 
have a value of $\theta$ near $90$ degrees. 
So if we only study the random variable 
$\theta$ for $\theta \le \theta_0$ with $\theta_0$ slightly less than $90$,  
i.e., condition on $\theta \le \theta_0$, we can reduce this problem. 
We take $\theta_0=85$. 

We can compute $b$ using eq. \ref{bformula} and the existing numerical
estimates of the exponents $\rho,\gamma$ and $\nu$. 
However, the result is not accurate enough for our purposes. 
So we use our data for the half-space hitting density to estimate $b$ 
as follows. Let $F_N(\theta)$ be the CDF when we use $N$ step SAW's, 
and let $H(\theta,b)$ be the predicted CDF. 
We assume that their difference is of the form
\bea
F_N(\theta) - H(\theta,b) \approx N^{-p} g(\theta) 
\eea
where the function $g(\theta)$ and the power $p$ are unknown. 
Let $\beta$ be an initial estimate of $b$ and write $b=\beta+\epsilon$.
So $\epsilon$ is small and we have
\bea
H(\theta,b) = H(\theta,\beta+\epsilon) \approx 
H(\theta,\beta) + \epsilon \frac{\partial H}{\partial \beta}(\theta,\beta)
\eea
So
\bea
F_N(\theta) - H(\theta,\beta) \approx N^{-p} g(\theta) +
\epsilon \frac{\partial H}{\partial \beta}(\theta,\beta)
\eea
We compute the left side from our simulation and we also compute
$\frac{\partial H}{\partial \beta}(\theta,\beta)$. 
We then use the data for the simulations with $N=100K, 250K, 500K$
to solve for $\epsilon, p$ and $g(\theta)$. 
We find $b=1.3303(3)$. The error bar given on this estimate is 
an educated guess obtained as follows. 
We find $p= 0.538$. When we rescale the differences
$F_N(\theta) - H(\theta,b)$ by $N^p$, the curves for different $N$ 
should collapse to a single curve. We obtain the error bar for $b$
by varying $b$ from our estimated value of $b=1.3303$ and seeing 
when this collapse clearly fails. A better method for studying 
the error in $b$ would be to run simulations for more values of $N$ 
and then use different sets of values of $N$ to estimate $b$. 


Figure \ref{fig_half} shows the difference between the simulation CDF
and the predicted CDF given by eq. \reff{cdf_half}
for the half-space for 
$N=100K, 250K, 500K$ and $1000K$.
One of the most important features in this figure is the vertical scale.
Even for the shortest length of $N=100K$, the maximum difference 
is only about $8 \times 10^{-4}$.
The error bars shown are plus or minus two standard deviations 
for the statistical errors in the simulation. 
They do not include the error from the finiteness of $N$. 
The figure shows that most of the difference comes from these 
finite $N$ errors, and these errors are going to zero 
as $N \ra \infty$. 

\begin{figure*}
\includegraphics{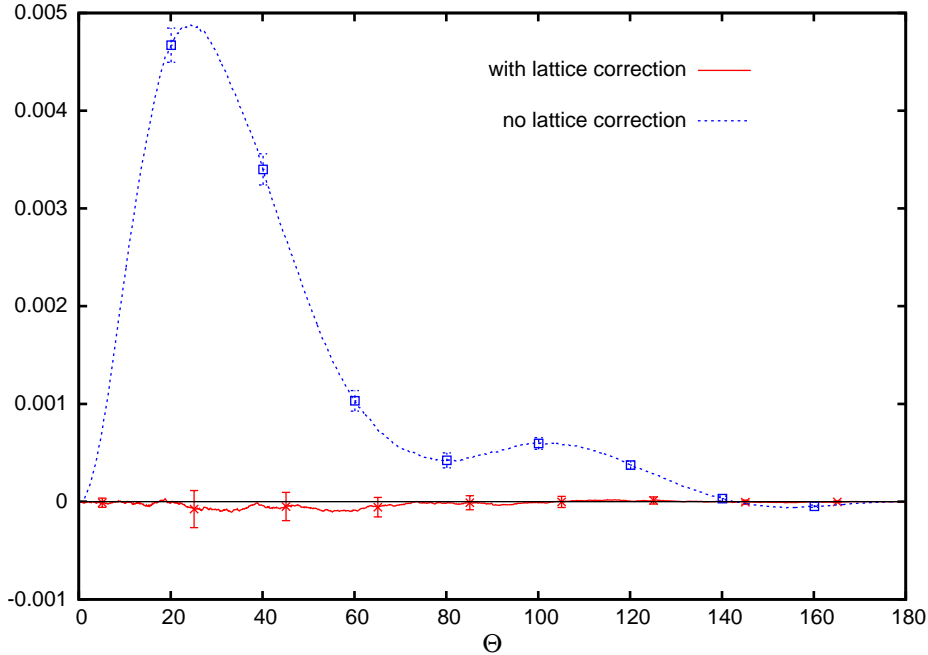}
\caption{
The curves are the difference between the simulation CDF's using 
$N=1000K$ for the hitting density for a sphere  centered at $(0,0,3/4)$
and the predicted CDF given by eq. \reff{cdf_sphere}. 
The curve that is nearly zero takes into account the lattice effect that
survives the scaling limit, and the other curve does not.
}
\label{fig_sphere}
\end{figure*}

The second prediction we test is 
the hitting density for the 
ensemble of SAW's in the sphere of radius $1$ centered at 
the origin which have one endpoint at 
$(0,0,a)$ and the other endpoint on the surface of the sphere. 
We take $a=3/4$.
As discussed in section \ref{sect_sphere} 
we use the fixed-length ensemble to study this hitting 
density. Given an $N$ step SAW in the full-space which starts at the origin,
we dilate the walk to produce a walk with one endpoint at the origin 
and the other endpoint on the sphere of radius $1$ centered at $(0,0,-a)$. 
We then condition on the event that 
this dilated walk lies entirely inside the sphere. 
We weight the walk by $R(\omega)^{-(\gamma-\rho)/\nu}$, where $R(\omega)$ is 
the dilation factor. As we argued in  section \ref{sect_half_space}
we expect the scaling limit to be the same as the 
scaling limit of the SAW in the sphere.
The lattice effects that were discussed in the previous section 
appear in this ensemble.
They persist in the scaling limit and 
must be taken into consideration when we test our prediction for 
the hitting density.
In figure \ref{fig_sphere}
we plot the difference of the predicted CDF given by eq. \reff{cdf_sphere}
and the CDF from the simulation for $N=1,000K$. Two curves are shown.
In one we take the lattice effect into account and in the other 
we do not. The figure shows that if we fail to account for the lattice
effect, then the difference between the predicted CDF and the simulation
CDF can be as large as $0.005$, much larger than the statistical errors
in our simulations and the finite-$N$ errors.

\begin{figure*}
\includegraphics{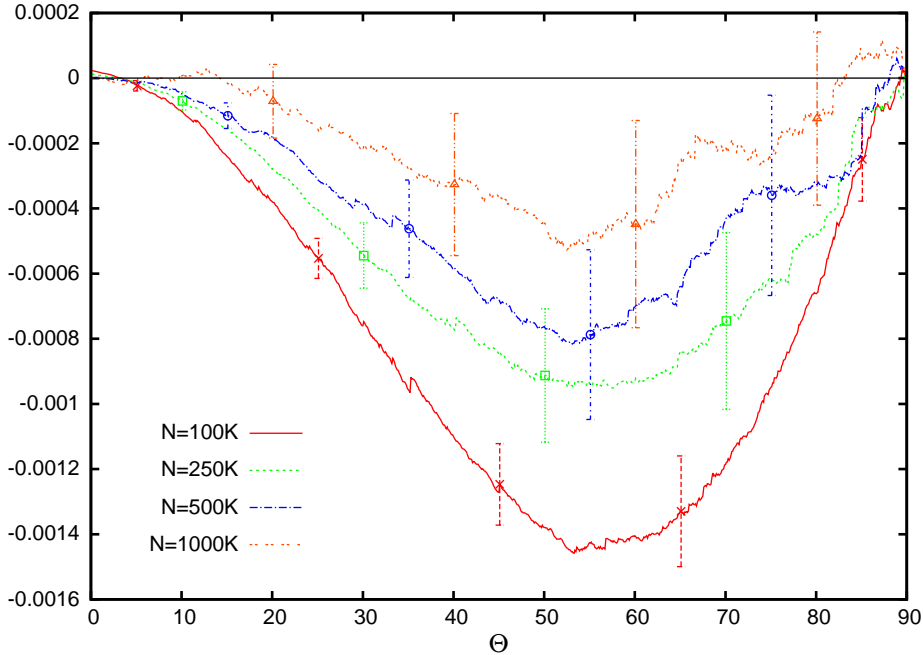}
\caption{
The curves are the difference between the simulation CDF's for 
the first hit of the bisecting plane for a SAW between two points
and the predicted CDF given by 
eq. \reff{cdf_bisect}. 
Four differences are shown corresponding to $N=100K, 250K, 500K, 1000K$. 
}
\label{fig_bisect}
\end{figure*}

The third prediction we test is for 
the ensemble of SAW's in the full-space 
from $(0,0,0)$ to $(0,0,2)$. Our prediction for this ensemble 
gives the distribution of the point on the plane $z=1$ where the 
walk first hits the plane.
We simulate this ensemble using the fixed-length ensemble.
This ensemble consists of SAW in the full-space with $N$ steps 
which start at the origin.
Given such a SAW we apply a Euclidean symmetry (rotation and dilation)
that fixes the origin and take the other endpoint of the SAW to $(0,0,2)$.
We weight the transformed $N$ step SAW $\omega$ 
by $||\omega(N)||^{-\gamma/\nu}$. As we argued in  section \ref{sect_pt_to_pt}
the scaling limit should be the same as the scaling limit of 
the ensemble of SAW's from $(0,0,0)$ to $(0,0,2)$.
In figure \ref{fig_bisect} we plot the differences between the 
simulation CDF and the predicted CDF given 
by eq. \reff{cdf_bisect} for $N=100K, 250K, 500K, 1000K$. 
Again the error bars are only for the statistical errors, and the 
figure shows the differences going to zero as $N \ra \infty$.

The pivot algorithm is a Markov chain Monte Carlo algorithm and so 
does not generate independent samples of the SAW.
For the hitting density for the half-space and the distribution
for the first hit of the bisecting plane, we sampled the Markov 
chain every 100 iterations. For $N=1000K$ we generated $10^9$ samples. 
For the smaller values of $N$ we generated on the order of 
$4 \times 10^9$ samples.
For the hitting density of the sphere we sampled the Markov chain 
every 10 iterations. However, in this simulation we must condition 
on the event that the dilated walk lies entirely in the sphere. 
The probability of this event goes to zero as $N \ra \infty$. 
For $N=1000K$ the probability is approximately $0.0025$. 
To compensate for this small probability we 
generated $60  \times 10^9$ samples.

%
%
%

\section{Conclusions}

We presented preliminary results on these tests of the conformal
invariance of the three-dimensional SAW in \cite{kennedy_3d_prl}.
The ensembles of SAW's for which there are predictions based on 
conformal invariance have walks with varying lengths. 
However, the fastest computational method for the SAW,  the pivot algorithm,
studies an ensemble of walks which all have the same number
of steps. A crucial tool in our study of these conformal invariance
predictions is the method explained in section \ref{sect_dilation}
of this paper
that allows us to study the ensembles that involve walks of varying 
lengths using the fixed-length ensemble. 

As in two dimensions, there are lattice effects that persist in the 
scaling limit that must be taken into account in our study 
of the prediction of the hitting density for the sphere. 
The computation of this lattice effect was discussed in section
\ref{sect_lat} of this paper.

For all three of our tests of the predictions of conformal invariance
the simulations presented in this paper 
show excellent agreement. With $N=1000K$ the maximum differences 
between the predicted CDF's and the CDF's found in the simulations 
range from $1 \times 10^{-4}$ to $6 \times 10^{-4}$.
The differences we see are due both to statistical errors
and the finite length of the walks we use. 
For two of the three predictions we have presented simulations 
for several values of $N$, the length of the SAW, which clearly
show the finite-$N$ error going to zero as $N \ra \infty$.

{\it Acknowledgments:} 
The support of the Mathematical Science Research Institute
in Berkeley, California where this 
research was begun is gratefully acknowledged.
An allocation of computer time from the UA Research Computing High Performance 
Computing (HPC) and High Throughput Computing (HTC) 
at the University of Arizona is gratefully acknowledged.

\end{document}